\documentclass[12pt,a4,epsf]{article}
\global\arraycolsep=2pt 
\input{epsf}
\usepackage{graphicx}
\begin{document}
\makeatletter
\def\fmslash{\@ifnextchar[{\fmsl@sh}{\fmsl@sh[0mu]}}
\def\fmsl@sh[#1]#2{%
  \mathchoice
    {\@fmsl@sh\displaystyle{#1}{#2}}%
    {\@fmsl@sh\textstyle{#1}{#2}}%
    {\@fmsl@sh\scriptstyle{#1}{#2}}%
    {\@fmsl@sh\scriptscriptstyle{#1}{#2}}}
\def\@fmsl@sh#1#2#3{\m@th\ooalign{$\hfil#1\mkern#2/\hfil$\crcr$#1#3$}}
\makeatother
\thispagestyle{empty}
\begin{titlepage}
\begin{flushright}
hep-ph/0107085 \\
LMU 01/10 \\
\today
\end{flushright}

\vspace{0.3cm}
\boldmath
\begin{center}
  \Large {\bf Calculation of the Higgs Boson Mass Using the
    Complementarity Principle}
\end{center}
\unboldmath
\vspace{0.8cm}

\unboldmath
\vspace{0.8cm}
\begin{center}
  {\large Xavier Calmet \footnote{supported by the Deutsche
Forschungsgemeinschaft, DFG-No. FR 412/27-1, \\
email:calmet@theorie.physik.uni-muenchen.de}}\\
  
\end{center}
\begin{center}
  and
  \end{center}
\begin{center}
{\large Harald Fritzsch}\\
 \end{center}
 \vspace{.3cm}
\begin{center}
{\sl Ludwig-Maximilians-University Munich, Sektion Physik}\\
{\sl Theresienstra{\ss}e 37, D-80333 Munich, Germany}
\end{center}
\vspace{\fill}
\begin{abstract}
\noindent
We compute the Higgs mass in a model for the electroweak interactions
based on a confining theory. This model is related to the standard
model by the complementarity principle. A dynamical effect due to the
large typical scale of the Higgs boson shifts its mass above that of
the W-bosons. We obtain $m_H=129.6$ GeV.
\end{abstract}
\end{titlepage}

Recently we have proposed a model for the electroweak interactions
based on a confining $SU(2)$ theory \cite{Calmet:2000th}. It was shown
that, at least at low energies, this model is complementary or dual to
the electroweak standard model \cite{Glashow:1961tr}. The
complementarity principle states that there is no phase transition
between the Higgs and the confinement phase if there is a Higgs boson
in the fundamental representation of the gauge group
\cite{Osterwalder:1978pc,'tHooft:1998pk}.  The Lagrangian of the
theory under consideration is exactly that of the standard model
before gauge symmetry breaking. However the sign of the Higgs boson
squared mass is changed, i.e. it is positive, and the gauge symmetry
is thus unbroken. We have the following fundamental left-handed
dual-quark doublets, which we denote as D-quarks:
\\
\begin{tabular}{lll}
leptonic D-quarks & $l_i=  \left(\begin{array}{c} l_1 \\ l_2 \end{array}

\right )$  &  (spin   $1/2$,  left-handed)  \\
& & \\
 hadronic  D-quarks   & $q_i= \left(\begin{array}{c}q_1 \\ q_2\end{array}
\right )$   &  (spin $1/2$, left-handed, $SU(3)_c$ triplet) \\
& & \\
scalar D-quarks  &  $h_i= \left(
  \begin{array}{c}
  h_1 \\ h_2
  \end{array}
\right )$ & (spin $0$). \\  \\
\end{tabular}
\\
The right-handed particles are those of the standard model.
As the gauge symmetry is unbroken, physical particles must be
singlets under $SU(2)$ transformation,  and we thus get the following
particle spectrum
\begin{eqnarray} \label{def2}
 \nu_L&=&\frac{1}{F}(\bar h l)=l_1+{\cal O}\left(\frac{1}{F}\right)
 \approx l_1 
  \nonumber \\
  e_L&=&\frac{1}{F}(\epsilon^{ij} h_i l_j)=
  l_2+ {\cal O}\left(\frac{1}{F}\right)  \approx l_2 
  \nonumber \\
   u_L&=&\frac{1}{F}(\bar h q)=q_1+{\cal O}\left(\frac{1}{F}\right) \approx q_1 
   \nonumber \\
     d_L&=&\frac{1}{F}(\epsilon^{ij} h_i q_j)=q_2
     +{\cal O}\left(\frac{1}{F}\right)\approx q_2
     \nonumber \\
     H&=&\frac{1}{2 F}(\bar h h)=h_{(1)}+\frac{F}{2} +
     {\cal O}\left(\frac{1}{2 F}\right)
     \approx h_{(1)}+\frac{F}{2} \\ \nonumber
     W^3_{\mu}&=& \frac{2 i}{g F^2} ( \bar h D_{\mu} h) = B^3_{\mu} +
+{\cal O}\left(\frac{2}{F}
   \right)  \approx B^3_{\mu}
    \nonumber \\
   W^-_{\mu}&=& \frac{\sqrt{2} i}{g F^2}
   ( \epsilon^{ij} h_i D_{\mu} h_j) = B^-_{\mu}+{\cal O}\left(\frac{2}{F}
   \right) \approx B^-_{\mu}
    \nonumber \\
   W^+_{\mu}&=& \left (\frac{\sqrt{2} i}{g F^2 } ( \epsilon^{ij} h_i D_{\mu} h_j)\right)^\dagger
 = B^+_{\mu}+{\cal O}\left(\frac{2}{F}
   \right)
   \approx
   B^+_{\mu},\nonumber   
\end{eqnarray}
where $g$ is the coupling constant of the gauge group $SU(2)_{L}$ and
$D_{\mu}$ is the corresponding covariant derivative. We have used the
unitary gauge
\begin{eqnarray}
   h_i=\left ( \begin{array}{c} F+h_{(1)} \\ 0 \end{array}\right),
\end{eqnarray}
where $F$ is the parameter appearing in the expansion of the bound
states (\ref{def2}). Matching the expansion of the Higgs boson to the
standard model we get $F=492$ GeV \cite{Calmet:2000th}. Using this
expansion, we can associate a certain scale, which is proportional to
$F$, to each particle. The scale of the $W$-bosons is then
$\Lambda_W=\sqrt{2}F/4=173.9$ GeV. As can be seen from the expansion
for the Higgs boson (\ref{def2}), there is a factor four between the
expansion parameter of the Higgs boson and that of the W-bosons, thus
one finds $\Lambda_H=\sqrt{2}F=695.8$ GeV. This factor four is
dictated by the algebraic structure of the underlying gauge theory.

In the confinement phase the Higgs boson is the $s$-wave of the
$SU(2)$ theory, whereas the $W$-bosons are the corresponding
$p$-waves. Thus one expects the Higgs boson to be lighter than the
$W$-bosons. But, as we shall show, a dynamical effect shifts the Higgs
boson mass above that of the $W$-bosons mass. The reason for this
phenomenon is the large Higgs boson scale compared to that of the
$W$-bosons.

\begin{figure}[h]
\begin{minipage}[t]{0.3\linewidth}\centering
\includegraphics[width=\linewidth]{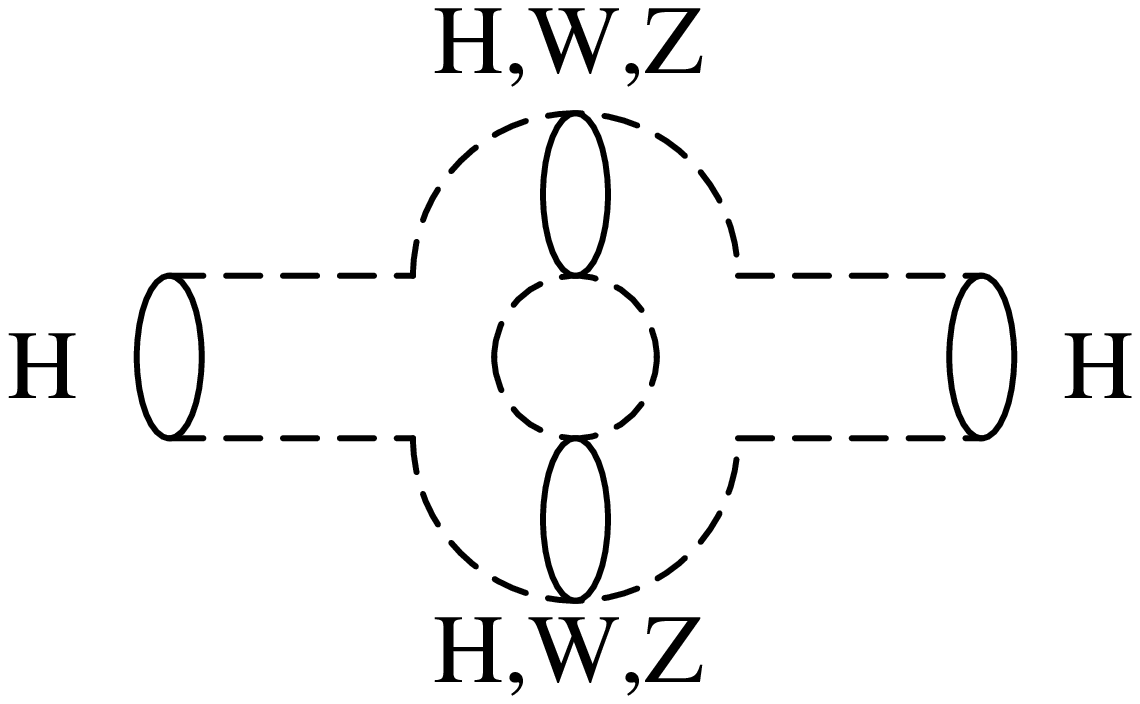}
\begin{minipage}{0.9\linewidth}
\caption{dual diagram: one loop 1PI contribution to $m_H$ \label{fig1}}
\end{minipage}
\end{minipage}
\begin{minipage}[t]{0.3\linewidth}\centering
\includegraphics[width=\linewidth]{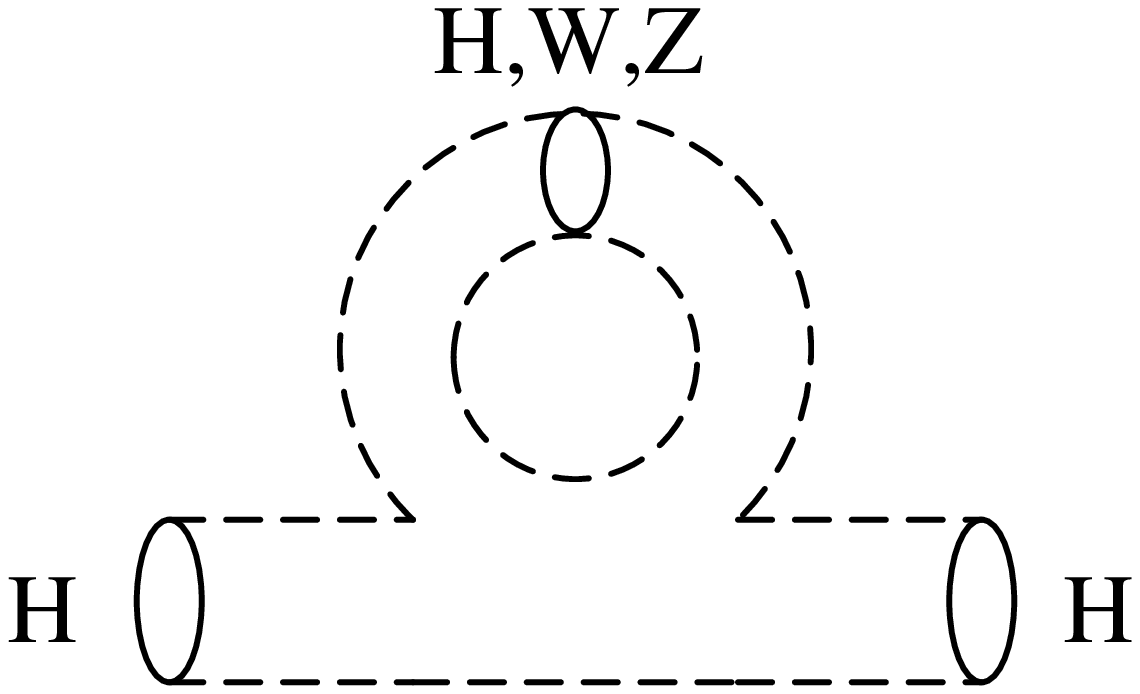}
\begin{minipage}{0.9\linewidth}
\caption{dual diagram: one loop 1PI contribution to $m_H$\label{fig2}}
\end{minipage}
\end{minipage}
\begin{minipage}[t]{0.3\linewidth}\centering
\includegraphics[width=\linewidth]{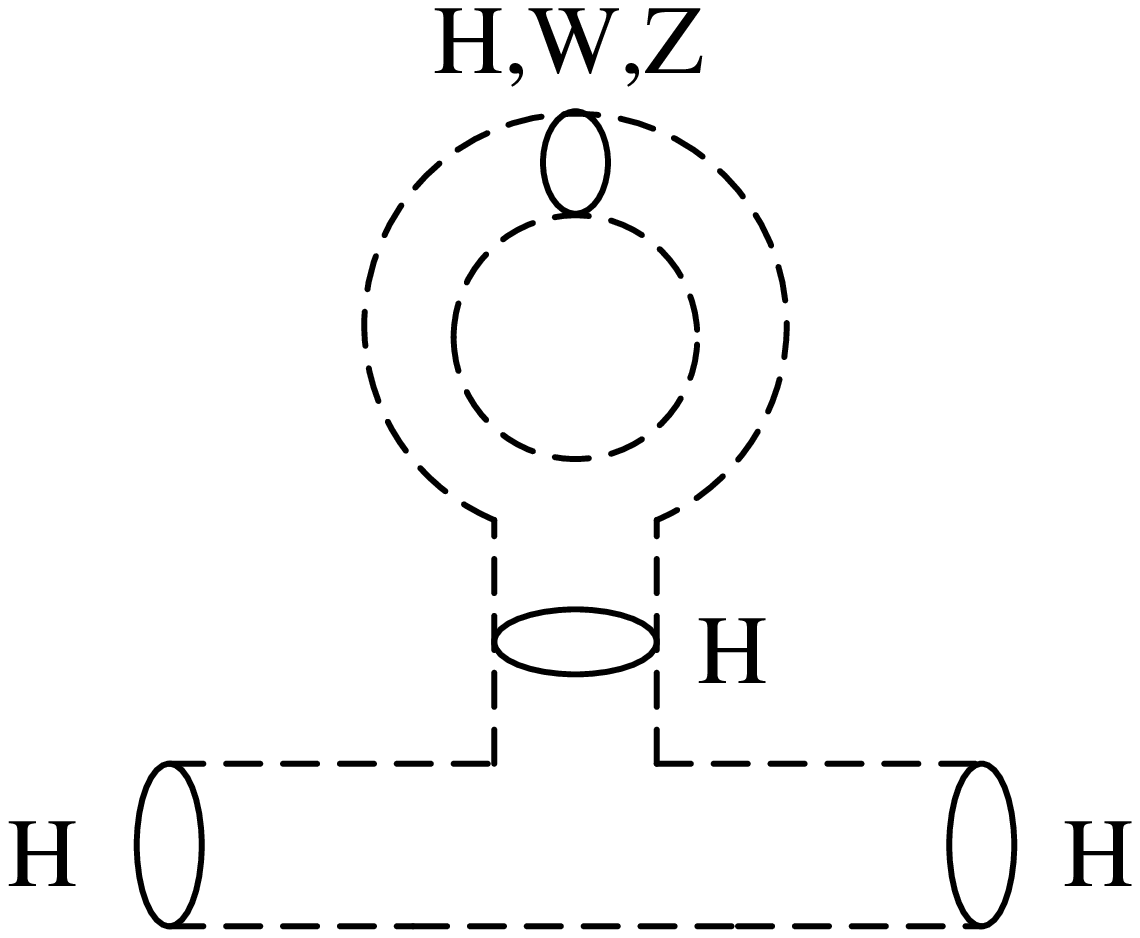}
\begin{minipage}{0.9\linewidth}
\caption{dual diagram: one loop 1PR contribution to $m_H$ \label{fig3}}
\end{minipage}
\end{minipage}
\end{figure}

The masses of the physical Higgs and W-bosons, being bound states
consist of a constituent mass $m_H^0=m^0_W=2 m_h$, where $m_h$ is the
mass of the scalar $D$-quark and of dynamical contributions. We have
to consider two types of diagrams: the one-particle reducible diagrams
(1PR) and the one-particle irreducible diagrams (1PI). For the Higgs
boson mass, we have to take the self-interaction and the contribution
of the $Z$ and $W^\pm$-bosons into account (see figures \ref{fig1},
\ref{fig2} and \ref{fig3}). The fermions couple via Yukawa coupling to
the Higgs boson, and as this interaction is not confining, fermions
cannot contribute to the dynamical mass of the Higgs boson.

The first task is to extract the constituent mass from the
experimentally measured $W$-bosons mass. The fermions contribute to
the dynamical mass of the $W$-bosons as they couple via $SU(2)$
couplings to the electroweak bosons but the divergence is only
logarithmic \cite{Veltman:1981mj} and we shall only keep the quadratic
divergences.
\begin{figure}[h]
\begin{minipage}[t]{0.3\linewidth}\centering
\includegraphics[width=\linewidth]{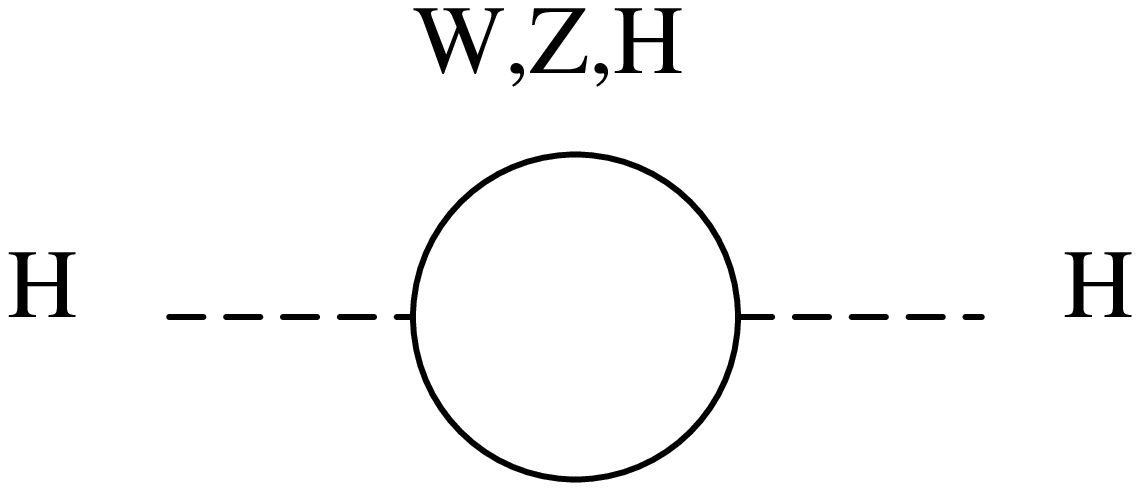}
\begin{minipage}{0.9\linewidth}
\caption{Feynman diagram: one loop 1PI contribution to $m_H$ \label{fig4}}
\end{minipage}
\end{minipage}
\begin{minipage}[t]{0.3\linewidth}\centering
\includegraphics[width=\linewidth]{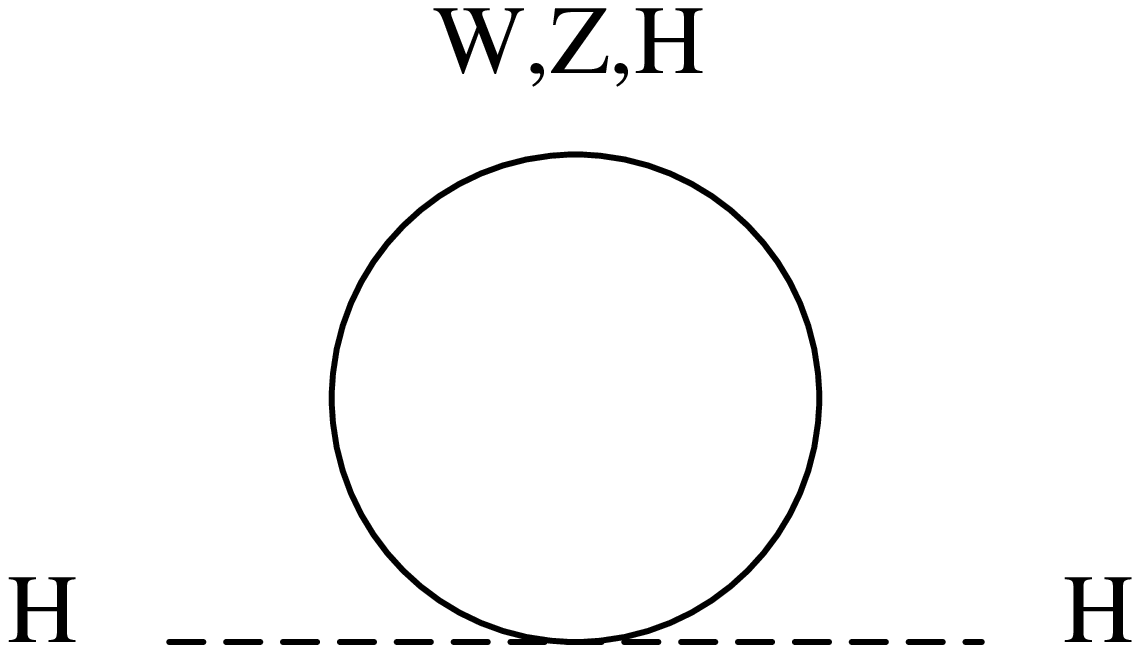}
\begin{minipage}{0.9\linewidth}
\caption{Feynman diagram: one loop 1PI contribution to $m_H$\label{fig5}}
\end{minipage}
\end{minipage}
\begin{minipage}[t]{0.3\linewidth}\centering
\includegraphics[width=\linewidth]{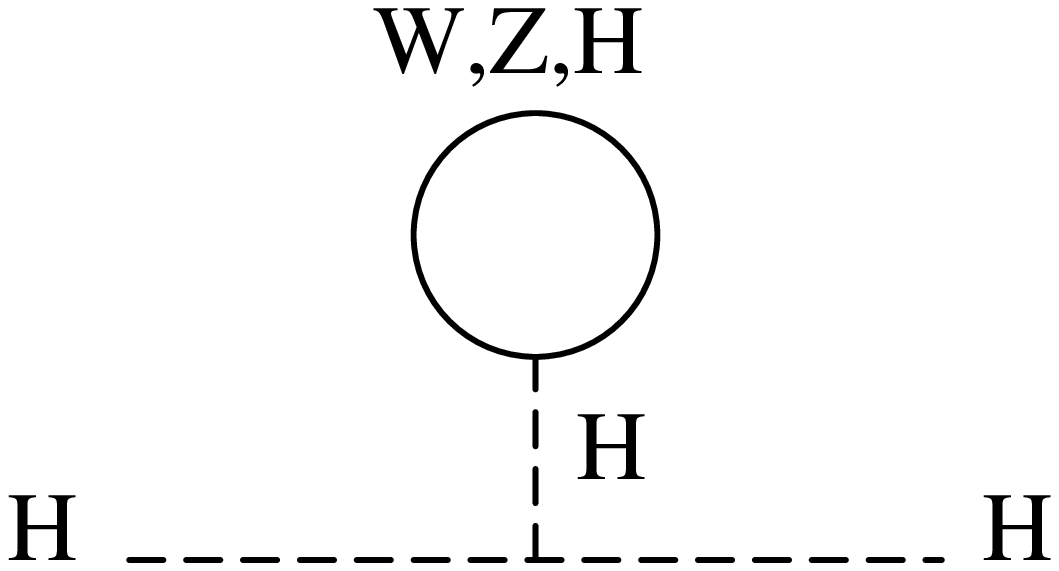}
\begin{minipage}{0.9\linewidth}
\caption{Feynman diagram: one loop 1PR contribution to $m_H$ \label{fig6}}
\end{minipage}
\end{minipage}
\end{figure}
We have considered the tadpoles and the one-particle-irreducible
contributions at the one loop order (the diagrams contributing to the
$W$-bosons mass are similar to those contributing to the Higgs-boson
mass).  Using the duality described in \cite{Calmet:2000th}, these
duality diagrams can be related to the Feynman graphs of figures
\ref{fig4}, \ref{fig5} and \ref{fig6}. The Feynman graphs have been
evaluated in ref.  \cite{Veltman:1981mj} as a function of a cut-off
parameter and we will only keep the dominant contribution which is
quadratically divergent.  We obtain:
\begin{eqnarray}
  m_W^2&=&{m_W^{0}}^2+\frac{3 g^2 \Lambda_W^2}{32 \pi^2 m_H^2}
  \left(m^2_H+2 m^2_W+m^2_Z \right).
\end{eqnarray}
This equation can be solved for $m_W^{0}$:
\begin{eqnarray}
 {m_W^{0}}^2&=&{m_H^{0}}^2= m_W^2-\frac{3 g^2 \Lambda_W^2}{32 \pi^2 m_H^2}
 \left(m^2_H+2 m^2_W+m^2_Z \right).
\end{eqnarray}

We can now compute the dynamical contribution to the Higgs boson mass.
The exact one loop, gauge invariant counterterm has been calculated in
refs. \cite{Veltman:1981mj}, \cite{Fleischer:1981ub} and
\cite{Ma:1993bt}. Using the results of ref. \cite{Ma:1993bt}, where
this counterterm was calculated as a function of a cut-off, we obtain:
\begin{eqnarray}
  m_H^2&=&{m_H^{0}}^2(m_H^2)+\frac{3 g^2 \Lambda_H^2}{32 \pi^2 m_W^2}
  \left(m^2_H+2 m^2_W+m^2_Z \right) \\ \nonumber &&
  + \frac{3 g^2 m_H^2}{64 \pi^2 m_W^2}
  \left(m_H^2 \ln \frac{\Lambda_H^2}{m_H^2} - 2 m_W^2 \ln \frac{\Lambda_H^2}{m_W^2} - m_Z^2 \ln \frac{\Lambda_H^2}{m_Z^2}\right).
\end{eqnarray}
The unknown of this equation is the Higgs boson's mass $m_H$. This
equation can be solved by numerical means. We obtain two positive
solutions: ${m_H}_1$=14.1 GeV and ${m_H}_2$=129.6 GeV. The first
solution yields an imaginary constituent mass and is thus also
discarded. The second solution is the physical Higgs boson mass. We thus
obtain ${m_H}$=129.6 GeV in the one loop approximation.  The
constituent mass is then $m_W^0$=78.8 GeV.

As expected the dynamical contribution to the $W$-bosons masses is
small and the Higgs boson mass is shifted above that of that of the
$W$-bosons mass because of the large Higgs boson scale.

Note that our prediction ${m_H}$=129.6 GeV is in good agreement with
the requirement of vacuum stability in the standard model which
requires the mass of the Higgs boson to be in the range 130 GeV to 180
GeV if the standard model is to be valid up to a high energy scale
\cite{Sher:1989mj}. We can thus deduce that the duality we have
described \cite{Calmet:2000th} must also be valid up to some high
energy scale. Our result is also in good agreement with the
expectation $m_H=98^{+58}_{-38}$ GeV based on electroweak fits
\cite{Kawamoto:2001ia}.

This has also consequences for the model proposed in
\cite{Calmet:2000vx}, where we assumed that the Higgs boson does not
couple to $b$-quarks.  Because of the dynamical effect we have
discussed, the Higgs boson is relatively heavy and should decay
predominantly into electroweak bosons.

\section*{Acknowledgements}
We should like to thank A. Hoang and A. Leike for useful discussions.

\end{document}